# Software Engineering Anti-patterns in Start-ups

E. Klotins, M. Unterkalmsteiner, T. Gorschek

Software start-up failures are often explained with poor business model, market issues, insufficient funding, or simply a bad product idea. However, inadequacies in software product engineering are relatively little explored and could be a significant contributing factor to high start-up failure rate.

In this paper we present analysis of 88 start-up experience reports. The analysis is presented in a form of three anti-patterns illustrating common symptoms, actual causes, and potential countermeasures of engineering inadequacies. The three anti-patterns are: product uncertainty comprising of issues in requirements engineering, poor product quality comprising of inadequacies in product quality, and team breakup comprising of team issues.

The anti-patterns show that challenges and failure scenarios that appear to be business or market related can actually originate from inadequacies in product engineering.

## 1 Introduction

Software start-ups are recognized as a source of innovative products. Cutting-edge software technologies enable start-ups to develop, and market software products fast with very limited resources. However, developing complex software products under uncertainty about markets, customer needs, technologies and with limited resources is challenging [1].

Different sources suggest that 75-99% of start-ups cease to exist within first few years of operation [1]–[3]. These failures can be explained with inadequacies in marketing, poor business model, lack of commitment or, simply, bad product ideas. However, inadequacies in engineering of software components of the product could have a substantial impact on the high start-up failure rate.

To explore to how software engineering is applied in start-ups and how inadequacies in software engineering are linked to start-up failures we examine 88 start-up experience reports. We apply qualitative analysis methods to identify recurring failure scenarios and their root causes. We present our results in a form of three anti-patterns: product uncertainty, poor product quality, and team breakup. The anti-patterns are depicted as cause-effect diagrams, showing the effect, symptoms of the effect and a root-cause analysis identifying potential origins of negative effects.

These anti-patterns are aimed to help practitioners to avoid potentially disastrous scenarios and to spot signs of such scenarios early. Moreover, along with the anti-patterns we discuss symptoms, shown on the left side of each figure, and potential countermeasures to break the anti-patterns on the right side of each figure.

# 2 Development of the anti-patterns

## 1.1 Data source

We use an online repository of start-up experience reports our data source [4], [5]. We screened the reports to remove reports from non-software start-ups or that were otherwise irrelevant. The final sample consisted of 88 reports with an average length of 1774 words.

The reports were written by principals of start-ups after a significant event, like acquisition, or going out of business. Contents of reports are not limited to software product engineering; the reports also discuss business, marketing and personal issues relevant to the start-up.

It is very important to note that our sample contains reports describing companies that are still operational, were acquired, and are out of business. Representation of operational and acquired companies is higher than average in the whole population.

## 2.1 Analysis method

We analysed each report and identified key analysis points, leading to gains or losses in software engineering or business development. Looking at the reporter's opinion on what activities had significant impact on software engineering, we attempt to identify factors contributing to the high impact situation.

We make use of descriptive (to summarize), process (to capture ongoing action) and evaluation (to assess the situation) coding jointly to capture analysis points in a experience report. Through analysis of the described situation we aim to differentiate between reported symptoms (e.g. running out of resources) and actual causes (e.g. poor resource planning due to lack of experience). Similar and recurring analysis points were grouped together to create the following anti-patterns.

# 3 Product uncertainty anti-pattern

Symptoms of product uncertainty are uncertainty of what new product features are required, uncertainty if customers are satisfied with already implemented features, and whether the product quality is on the desired level. A further implication of this anti-pattern is poor product performance in the market.

The Product uncertainty anti-pattern (Figure 1) is comprised of inadequacies in requirements engineering. Further consequences of product uncertainty are wasted resources by developing unwanted features and poor results in marketing the product.

The key causes for product uncertainty are a vague formulation of the product's benefits and neglect of user feedback. The uncertainty stems from the inability to articulate engineering decisions with actual user needs.

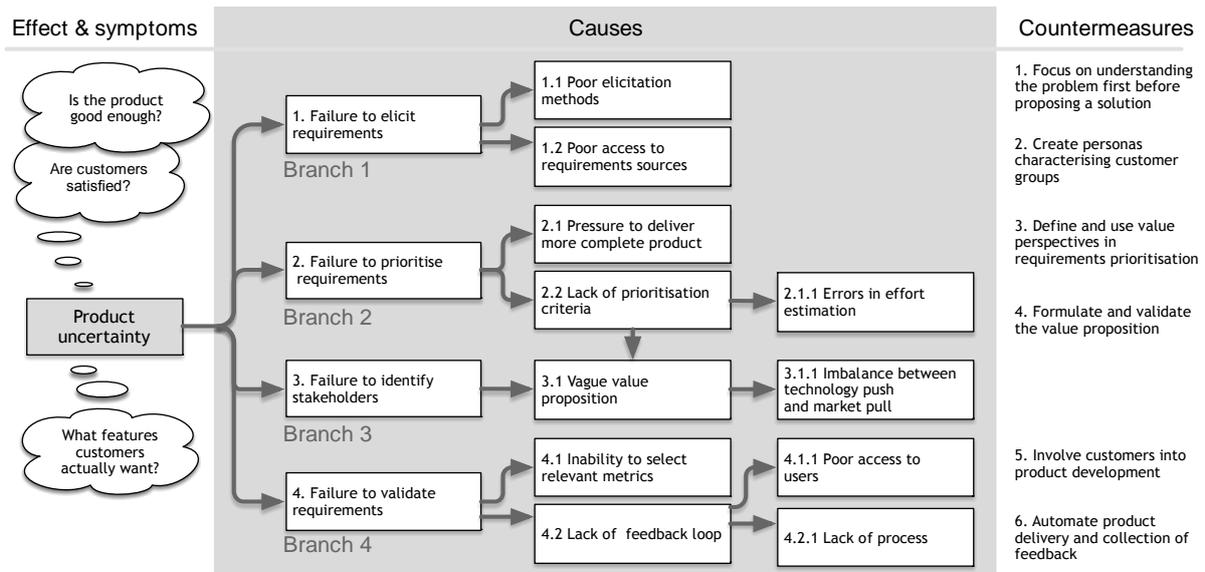

*Figure 1, Engineering uncertainty anti-pattern*

## 3.1 Analysis of the causes and potential countermeasures

Requirements elicitation, (Branch #1 in Figure 1), is a process of extracting requirements from their source. A common source of requirements are potential and existing customers. Interviews are reported as a common way to gather customer feedback. However, quality of the interview data depends on the interviewer's skill to ask the right questions and to select the right interviewees.

A common pitfall is to mistake interested people for potential customers. Initially, founders may solicit feedback on their product idea from family members or friends. However, these people may show interest and support the idea only just because they are curious and trying to be kind, and not because they see themselves as potential customers. As emphasized by one start-up "*You must never, ever, pitch the product to the customer and ask for their feedback*".

The initial focus of requirements engineering should be on understanding a customer problem and the actual customer behaviour. The interviewer should first understand customers' perspective on the problem. As *Dinnr*, a grocery delivery start-up puts it: "*Do not ask about customer's future intentions, but ask about their past behaviour*". Superimposing a solution and asking for feedback with questions like "Do you think my idea X is good for Y?", begs for positive feedback and provides a misleading input.

Start-ups get their ideas for product features from brainstorming, similar products and customer feedback. However, not all ideas are feasible and equally good. Selecting only most relevant ideas for implementation is a task for *requirements prioritization* (Branch #2). A common cause for difficulties in requirements prioritization is absence of clear prioritization criteria.

*Prioritization criteria* stem from the product value proposition and are used to rank different ideas. Commonly used criteria are value and effort. Features that require the least effort and provide the most value are implemented first. However, if the value proposition is vague it could be difficult to identify the most important features. As described by *Disruptive Media*: *"We kept building more features, since we always felt that the service needs X because Flickr has it too or he/she said he needs that feature"*.

Different people may have different views on what is valuable, therefore making prioritization complicated. There are also other factors at play when deciding whether to implement a feature or not. For example, availability of skill and technology, market potential, relevance to the product value proposition, demand from an important stakeholder and so on. Balancing these aspects is a challenge. Failure to prioritize the right features leads to scope creep, wasted resources and missed market opportunities.

A potential countermeasure is to recognise different perspectives on value. For example, there could be customer, business, sales and engineering perspectives. When negotiating value of a feature, each perspective can be considered separately and their compound value used to characterize the total value of the feature. For example, a feature may deliver a lot of customer value but also contribute negatively to engineering by introducing technical debt. Thus, a total value of such feature may be lower than expected. Ideally, product features are selected for implementation in a way that balances all perspectives, thus increasing the total value of the product.

Identification and access to potential customers is essential to elicit product requirements and to perform requirements validation. The first customers who had participated in product development are likely to become the first paying customers of the product. A common cause for difficulty to identify potential customers is a vague value proposition. Having a too broad definition of the targeted customers is likely to cause difficulties in identifying what exact value the product is providing.

*A value proposition* is a short summary of the promised product benefits, to whom these benefits are relevant, and competitive advantages of the product [6]. If the promised benefits are vague it could be difficult to identify a distinct set of potential customers. Too much focus on possibilities of new technologies instead of actual customer needs, could be a cause for vague value proposition. As described in one report *"An start-up needs to showcase its product, to demonstrate its benefits and the value added to any customer willing to pay for it."* Moreover, the value proposition must be validated and continuously refined with potential customers.

The process of formulating a value proposition is not a software engineering task per se. However, creating personas, fictional characters representing different customer groups, could help to spot too vague formulations of the product customers. The process of creating personas allows to discover new requirements [7].

In start-ups products are often innovative and unknown to both customers and developers. Therefore, it is important to continuously check if developers have understood the customer problem correctly and the envisioned solution is useful to customers, i.e. to perform requirements validation (Branch #4). A common hindrance to requirements validation is lack of a feedback loop.

*The Feedback loop* is a process to present an artefact, e.g. a product prototype, to customers, collect their feedback and to use this feedback to improve the artefact. Then, repeat the demonstration, collect the feedback and improve until a desired level of quality and functionality is achieved.

Not using customer feedback creates a risk on wasting resources on building features that does not provide enough value to customers. As described by one practitioner: *"We rarely had meaningful conversations with our target end-users. We huddled together to decide on ideas that sounded nice, built prototypes, put on our salesman hats, and didn't understand why we weren't closing deals"*.

To establish the feedback loop a product *development process,* involving potential customers and embracing their feedback is required. Agile practices aim to support customer involvement in product development and proposes practices such as iterations, documenting requirements in natural language using user stories, and backlog grooming to make sure that the requirements are up to date. Use of such practices depends on access to customers.

Access to customers could be hindered by inadequacies in stakeholder identification (no meaningful group of customers is identified) and a distance between the developers and the customers (getting in touch with customers is difficult). Speed of product development depends on how fast the feedback loop could be executed. As *Disruptive Media* described the difficulties to iterate fast: *"Small print labs were often quite busy and did not have time to immediately have a look at the new version and give feedback"*. Nevertheless, speed can be improved by releasing products more frequently, using automated build and deployment tools, and automated collection of user feedback.

## 4   Poor product quality anti-pattern

Symptoms of the poor product quality anti-pattern are insufficient and degrading product quality. External symptoms are, for example, failures in production, lower than acceptable usability and reliability, that stems from shortcomings in the internal quality. Symptoms of poor internal quality are ever increasing difficulty to make any changes in the product due to accumulation of technical debt. This anti-pattern comprises of inadequacies in quality requirements management and software architecture (see Figure 2). An immediate consequence of this anti-pattern is damaged reputation through poor product reviews. As in a case of *Flowtab*, *"A big-bang product launch involving many partners and customers become a catastrophe when the app failed"*. Further consequences are long-term effects of technical debt, such as expensive product maintenance, difficult to fix quality issues and decaying team morale.

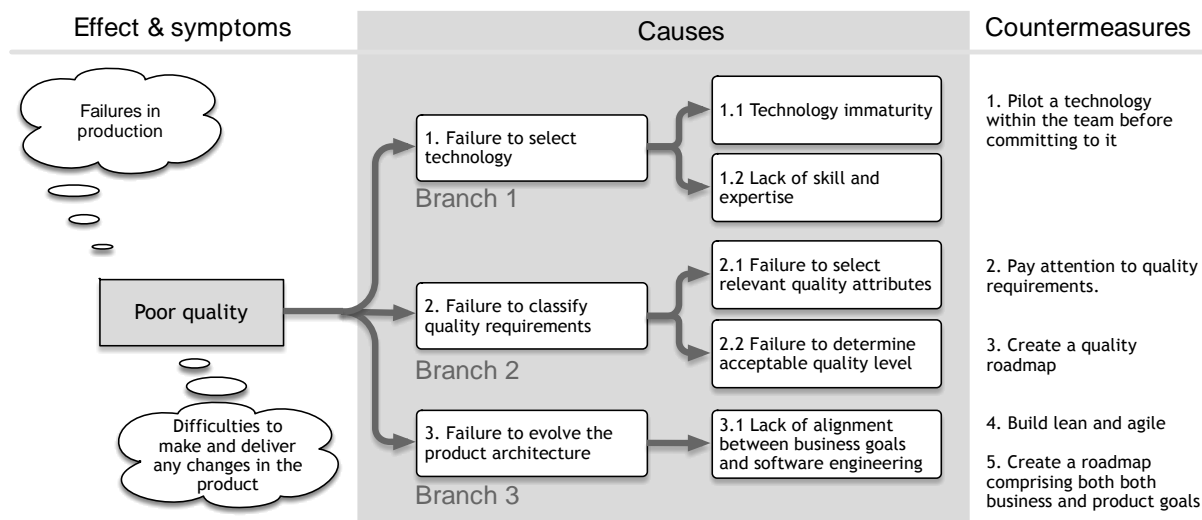

*Figure 2, Poor product quality anti-pattern*

## 4.1 Analysis of the causes and potential countermeasures

Start-ups often leverage on cutting-edge technologies to support their innovation endeavours. However, new technologies are often immature, lacking robustness and documentation. *Seismic*, a start-up aiming to enter early mobile video market reflects on this: "*We had to use Adobe Flash on a browser which was terrible in so many ways I can't even start to explain how bad it was*". Thus, achieving a competitive quality could be challenging. More importantly, there could be a shortage of experts on a new technology, leading to sub-optimal solutions, and a prolonged learning curve.

*Development technologies* are tools, programming languages, third-party components and other elements essential to compose the product. When selecting development technologies, engineers need to balance promised benefits, limitations, specific software requirements, availability of skills and other factors. The difficulty lies in uncertainty about true benefits of using a cutting-edge technology, future needs, and potentially unforeseen pitfalls of a particular technology. That said, less experienced developers could make sub-optimal decisions even when working with a robust technology. As stated in the report by *Parceld*, their team lacked essential development competencies: "*My team and I started using Dropbox before switching to Github, and it was too late for some of the code*."

Before committing resources to a unfamiliar technology, it is useful to try out the technology on a small-scale project [8]. A pilot project could provide insights whether the team skills are appropriate, help to estimate the learning curve, assess whether the technology fits its purpose, and to provide a reference point for effort estimation.

Due to resource limitations, a start-up must prioritize which quality attributes are relevant for the product and what is the optimal level of each attribute. As shown in Branch #2, a potential cause for poor quality is inadequacies in handling quality requirements.

Quality requirements determine how a product should work, rather than what a product does. For example, how fast the product should respond to user input, or how easy to use the product should be. Quality considerations drive decisions behind the product architecture.

Due to resource limitations and time pressure it is impractical to aim for best possible quality level for all quality aspects. Rather, a start-up can focus only on maximising a few key quality aspects and aim for an acceptable quality level for all other aspects. As described by one start-up: *"We may have missed opportunities, or not gotten the minimum quality that we needed because we were so focused on capital preservation."* Inadequacies in determining the key quality attributes stem from inadequacies in requirements engineering, as illustrated in the previous anti-pattern.

A potential solution to tackle quality challenges is to create a quality roadmap. Regnell et al. [9] suggests to determine four quality level breakpoint characteristics for a product domain and use them to guide any decision on quality targets. The proposed quality breakpoints are "useless", "useful", "competitive" and "excessive". Determining the breakpoints enables to set meaningful quality targets for upcoming product releases.

Another potential cause for poor quality stems from failure to evolve the product architecture to changing business needs (Branch #3). Software architecture is a metaphor, referring to how different components of the product are combined to create desired functionality and quality, similar to the architecture of a building.

A typical scenario for a start-up is to start by developing a makeshift product prototype to collect feedback and to test out the business idea. Then, over time, the prototype evolves into a fully-fledged product ready to provide a robust and scalable solution. As the goals shift, the architecture must evolve as well to support the new requirements. A common pitfall is to invest too much into architecture upfront, before a need for such architecture is validated. Such upfront investment could turn to waste if the features the architecture is aimed to support turn out to be irrelevant. As in the case of *Saaspire*: "*The custom platform created a massive overhead on our development work. We ended up over engineering our systems so they could support both today's and tomorrow's products*."

Another pitfall is to neglect the product architecture and patch it only when a failure occurs. In this scenario, the product accumulates technical debt and the overall quality of the product erodes until the effort of maintaining the product becomes too high to be viable.

Agile architecture practices can help to create a robust architecture without an upfront investment and in uncertain conditions. Woods [10] suggests to make a minimum number of key decisions at the beginning, leverage on architecture patterns, deliver the architecture incrementally, and resolve cross-cutting concerns, such as security and scalability, with clear and proven solutions.

# 5 Team breakup anti-pattern

Symptoms of this anti-pattern are key people leaving the company and faultlines in the team. The effect of this anti-pattern is a loss of key resources hindering the team's performance.

Moreover, the effects of this anti-pattern amplifies its root causes. If a critical person, a founder or person providing unique skills, decides to leave the company it could cause a turmoil breaking up the team.

The team breakup anti-pattern comprises of issues in team management (see Figure 3). Breakup of a team could have potentially disastrous and irreversible effect on star-up's future prospects. With key human resources and knowledge gone it could be very difficult to resume normal operations. As experienced by a principal of *Manilla*: "*As I thought things couldn't get any worse, our CTO and cofounder decided to quit. Development speed would decrease dramatically and the driving force in getting stuff done would be gone*".

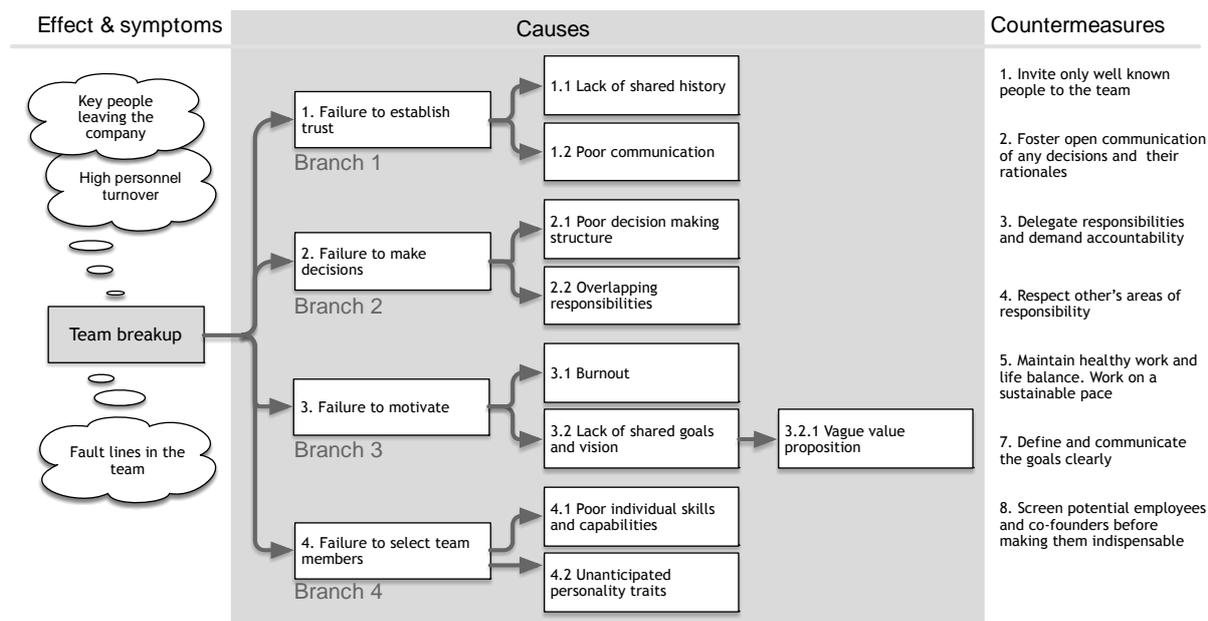

*Figure 3, Team breakup anti-pattern*

## 5.1  Analysis of the causes and potential countermeasures

Lack of trust among team members is a likely contributing factor to team disintegration (Branch #1). Team of *Dijiwan* describes trust as: "*Trusting is accepting. Successes and mistakes.*" Typical causes for low mutual trust are poor communication and lack of shared history.

When a level of trust in the team is low, each member acts to protect own interests' first. This leads to sub-optimal performance, and unwillingness to give away control. Trust can be improved by openly communicating any relevant objectives, communicating a rationale behind decisions, delegating responsibilities, and seeking accountability.

Another possible cause for team breakup stems from inadequacies in decision making, potentially caused by overlapping responsibilities and unclear decision making structures (Branch #2). To be able to make own decisions, employees need to be aware of the limits of their own and others authority.

Start-up teams constantly face decisions, some may have major impact and some decisions are minor with little consequences. Regardless of the decision at hand, it is important that the decision-making process is efficient, transparent and leaves all stakeholders satisfied. The experience reports discuss a number of practical issues in decision making. For example, the founders micromanaging trivial decisions rather than delegating them creating a bottleneck; profound disagreements between founders leaving important decisions in a deadlock; and team members violating other's areas of responsibility. These scenarios leave some or all stakeholders dissatisfied thus contributing to breakup of the team.

As a countermeasure, areas of responsibility need to be defined, communicated and respected. This enable accountability of any outcomes from a decision, improves efficiency of decision making and reduces friction in the team.

Lack of motivation within the team could be another cause for a team breakup (see Branch #3). Likely causes for a degrading motivation are lack of clear and shared objectives, and a burnout from constant stress.

Stress caused by never ending struggles with launching a product, and poor work/life balance could lead to a burnout, loss of motivation, and harm one's physical health. A countermeasure is to work on a sustainable pace and to keep a healthy work/life balance. As one start-up founder described it: "*I felt really burned out and demotivated. I realized that my start-up just didn't fit with my personal goals anymore and left me physically and emotionally exhausted*".

Absence of clear and shared objectives could be another cause for degrading motivation. It could be an issue in communicating the objectives within the team and engaging the team to attain the objectives. However, as suggested by the reports, a potential cause could also be vaguely formulated overall objectives of the start-up. A potential countermeasure is to make sure that everyone in the team knows and is committed to a common goal.

The team and its capabilities is one of the main assets in a start-up company. Failure to select team members with adequate skills and matching personality types could be a potential cause for team breakup (Branch #4).

Adequate levels of individual skills is a prerequisite for efficient collaboration in a team. Individual skills are divided into "hard" skills, such as proficiency in a programming languages, and "soft" skills, such as emotional intelligence, teamwork and communication skills.

Poor individual skills may hinder the overall team's performance. However, the reports suggest that insufficient "hard" skills can be balanced with a willingness to learn and a good environment for learning. As suggested by *Shnergle*, a start-up building a location based image service: "*Hire for character and attitude, flexibly scale up cost (salary) with competence. Our developers learned fast, worked crazy hard, and put together a decent product with little previous experience*".

Personality traits determine how a person behaves under specific circumstances. Psychology studies identify a number of distinct traits as dimensions**.** For example, a person may be more social or shy, more open to new ideas or rather conservative.

Unanticipated personality traits may remain hidden until triggered and could contribute to a team breakup. For example, a key person could go rogue and start pushing goals that are beneficiary to him personally, and not the team as a whole. As experienced first-hand by a co-founder of Dijiwan: "*Unjustified expenses by our CEO were burning our cash fast. I never had access to the accountancy results. It bothered me but I did nothing because I never thought it could possibly be that bad, that fast*.". Even when the person is removed, damage is already done.

A potential countermeasure, as suggested by *Boompa*, a start-up building a social network for car enthusiasts, is to invite only previously known people to the start-up's core team. In their own words: "*It's why we decided to work as a duo, even though we had enough money for more. The trust just wasn't there with anyone else. You can hire employees anytime, but you only get to pick your partners once.*" People with shared history are more likely to be aware of each other's skills and characters.

# 6 Conclusion

The analysis of the start-up experience reports revealed that some failures that appear to be marketing or business related could actually be rooted in inadequacies of product engineering. For example, while poor sales results could be explained with sub-optimal marketing & sales strategy, they may stem from a product not solving an actual customer problem, hinting towards poor requirements engineering. If inadequacies in requirements engineering are realized too late, a start-up may not have enough resources to go back and rethink their product.

Another example is that a start-up may struggle with resource shortages for developing new features and attribute that to insufficient funding. However, the struggle may originate from sub-optimal product architecture and accumulation of technical debt.

Looking at the evidence from the reports, most start-ups experience all anti-patterns to some extent. Start-ups experience a degree of product uncertainty, make sub-optimal decisions, and experience team issues. What makes the difference is whether these inadequacies are realized and mitigated early, before their compound effect turns out to be disastrous to the company.

# 7 References

[1]     S. Blank, "Why the Lean Start Up Changes Everything," *Harv. Bus. Rev.*, vol. 91, no. 5, p. 64, 2013.

[2]     I. PitchBook Data, "U.S. Middle market report Q4 2015," 2015.

[3]     I. PitchBook Data, "European Middle Market Report 2H 2015," 2015.

[4]     E. Klotins, M. Unterkalmsteiner, and T. Gorschek, "Supplementary material, 88 start-up experience reports." [Online]. Available: http://eriksklotins.lv/files/exp-reports-study-supplemental-material.pdf. [Accessed: 15-Jun-2017].

[5] "CB Insights." [Online]. Available: https://www.cbinsights.com/blog/startup-failure-post-mortem/.

[6] C. R. Carlson and W. W. Wilmot, *Innovation: The five disciplines for creating what customers want*. Crown Business, 2006.

[7] J. Pruitt and J. Grundin, "Personas : Practice and Theory," *Proc. 2003 Conf. Des. user Exp.*, pp. 1–15, 2003.

[8] G. C. Murphy, I. C. Society, R. J. Walker, S. Member, and E. L. a Baniassad, "Technologies : Lessons Learned from Assessing Aspect-Oriented Programming," vol. 25, no. 4, pp. 438–455, 1999.

[9] B. Regnell, R. B. Svensson, and T. Olsson, "Supporting roadmapping of quality requirements," *IEEE Softw.*, vol. 25, pp. 42–47, 2008.

[10] E. Woods, "Aligning Architecture Work with Agile Teams," *IEEE Softw.*, vol. 32, no. 5, pp. 24–26, 2015.